\documentclass[%
 superscriptaddress,
 preprint,
 preprintnumbers,
 aps,
 prb,
showkeys]{revtex4-2}

\usepackage{booktabs,subcaption,amsfonts,dcolumn,makebox} 
\usepackage[math]{cellspace}
\usepackage{makecell}
\usepackage{cellspace}
\usepackage{graphicx}
\usepackage{etoolbox}                
\AtBeginEnvironment{tabular}{\small} 
\usepackage[dvipsnames]{xcolor}

\begin{document}
\newcommand\mc[1]{\multicolumn{1}{c}{#1}}


\title{Temperature dependence of the Raman spectrum of orthorhombic Bi$_2$Se$_3$}


\author{Irene Mediavilla-Martínez}
\affiliation{GdS-Optronlab
group, Dept. Física de la Materia Condensada, Cristalografía y Mineralogía,
Universidad de Valladolid, Paseo de Belén 19, 47011 Valladolid, Spain}

\author{Christian Kramberger}
\affiliation{Faculty of Physics, University of Vienna, Vienna, Austria}
\author{Shabnam Dadgostar}
\affiliation{GdS-Optronlab
group, Dept. Física de la Materia Condensada, Cristalografía y Mineralogía,
Universidad de Valladolid, Paseo de Belén 19, 47011 Valladolid, Spain}
\author{Juan Jiménez}
\affiliation{GdS-Optronlab
group, Dept. Física de la Materia Condensada, Cristalografía y Mineralogía,
Universidad de Valladolid, Paseo de Belén 19, 47011 Valladolid, Spain}
\author{Paola Ayala}
\affiliation{Faculty of Physics, University of Vienna, Vienna, Austria}
\author{Thomas Pichler}
\affiliation{Faculty of Physics, University of Vienna, Vienna, Austria}
\author{Francisco J. Manjón}
\affiliation{Instituto
de Diseño para la Fabricación y Producción Automatizada, MALTA-Consolider Team,
Universitat Politècnica de València, 46022 Valencia, Spain}
\author{Plácida Rodríguez-Hernández}
\affiliation{Departamento de Física, Instituto de Materiales y Nanotecnología, Universidad de La Laguna, 38205 Tenerife, Spain}
\author{Alfonso Muñoz}
\affiliation{Departamento de Física, Instituto de Materiales y Nanotecnología, Universidad de La Laguna, 38205 Tenerife, Spain}
\author{Nadezhda Serebryanaya}
\affiliation{Technological Institute for Superhard and Novel Carbon Materials,  Centralnaya Street 7a, 108840 Troitsk, Moscow, Russian Federation}
\author{Sergei Buga}
\affiliation{Technological Institute for Superhard and Novel Carbon Materials,  Centralnaya Street 7a, 108840 Troitsk, Moscow, Russian Federation}
\affiliation{Moscow
Institute of Physics and Technology Institutskiy per. 9, 141700 Dolgoprudny,
Moscow Region, Russian Federation}
\author{Jorge Serrano}
\email[Corresponding author: ]{jorge.serranogutierrez@gmail.com}
\affiliation{GdS-Optronlab
group, Dept. Física de la Materia Condensada, Cristalografía y Mineralogía,
Universidad de Valladolid, Paseo de Belén 19, 47011 Valladolid, Spain}
\affiliation{Meeting Life LLC, 1210 Washington Ave. Ste. 213, Miami Beach, Florida 33139, USA}



\date{\today}

\begin{abstract}
Bismuth selenide, a benchmark topological insulator, grows in a trigonal structure at ambient conditions and exhibits a number of enticing properties related to the formation of Dirac surface states.
Besides this polytype, a metastable orthorhombic modification with {\em Pnma} space group, o-Bi$_2$Se$_3$, has been produced by electrodeposition and high-pressure high-temperature synthesis. It displays significant thermoelectric properties in the mid-temperature range, particularly upon Sb doping. 
However, very little experimental information is available on the fundamental properties of this orthorhombic polytype, such as, e.g., the electronic band gap and the lattice dynamics.

We report here the temperature dependence of the Raman spectra of o-Bi$_2$Se$_3$ between 10~K and 300~K, which displays an anharmonic behavior of the optical phonons that can be modelled with a two-phonon decay channel. 
In order to analyze the data we performed {\em ab initio} calculations of the electronic bandstructure, the phonon frequencies at the center of the Brillouin zone, and the phonon dispersion relations along the main symmetry directions, examining the effect of spin-orbit coupling in both phonon and electronic energies.  
Lastly, we report here cathodoluminescence experiments at 83~K that set a lower limit to the electronic bandgap at 0.835~eV, pointing to an indirect nature, in agreement with our calculations.
These results shed light to essential properties of orthorhombic Bi$_2$Se$_3$ for further understanding of the potential of this semiconductor for thermoelectrics and other relevant applications.
\end{abstract}


\maketitle

\newpage
\section{Introduction\label{sec:intro}}
The quest for advanced materials with tailored applications has been paved with the discovery of fascinating properties, sometimes unexpectedly.
One of the benchmark examples of this combination of new physics and enticing applications is the case of bismuth selenide, with a trigonal (R$\bar{3}$m) tetradymite structure, $\alpha$-Bi$_2$Se$_3$, a three-dimensional topological insulator, i.e., a material characterized by an insulating bulk and conducting surface states. With a single Dirac cone on the Fermi surface~\cite{Zhang2009}, $\alpha$-Bi$_2$Se$_3$ exhibits spin-momentum locking of massless helical Dirac fermions~\cite{Hsieh2009}, due to strong spin-orbit coupling (SOC), coexisting with a two-dimensional electron gas~\cite{Bianchi2010}, and Landau quantization of the surface states~\cite{Cheng2010}.  
Trigonal Bi$_2$Se$_3$ has been suggested to exhibit superconducting behavior, demonstrated with Cu doping~\cite{Hor2010}. In nanoribbons it presents pronounced Aharonov-Bohm oscillations in magnetoresistance~\cite{Peng2009} that point to potential applications in future spintronic devices at room temperature. More recently, charge-curren$\alpha$-induced spin polarization has been detected electrically in $\alpha$-Bi$_2$Se$_3$ films~\cite{Li2014} due to locking of spin-momentum using a ferromagnetic metal tunnel barrier surface contact to measure voltage. This opens a pathway to control spin state population for potential applications in quantum computing. Moreover, $\alpha$-Bi$_2$Se$_3$ films have been shown to generate a spin-transfer torque on adjacent ferromagnetic thin films~\cite{Mellnik2014}, implying that topological insulators could facilitate efficient electrical manipulation of magnetic materials at room temperature, for example in logic and memory applications. Coupling $\alpha$-Bi$_2$Se$_3$ with a ferromagnetic insulator in a bilayer system, such as EuS, has also been shown to demonstrate topologically enhanced ferromagnetism persisting up to room temperature, due to proximity coupling~\cite{Katmis2016}, highlighting further the potential for these applications. 
High-performance large-area electronics is another arena where $\alpha$-Bi$_2$Se$_3$ has been proposed for applications, together with 2D materials such as WSe$_2$, NbSe$_2$, In$_2$Se$_3$, Sb$_2$Te$_3$, and black phosphorus~\cite{Lin2018}.

Despite this broad range of applications, thermoelectric properties are much less appealing in $\alpha$-Bi$_2$Se$_3$ than in its heavier analogue, Bi$_2$Te$_3$. With similar trigonal structure, Bi$_2$Te$_3$ is known to display superior thermoelectric properties at room temperature due to the presence of six valleys for the highest valence band in the electronic bandstructure, versus only one available in $\alpha$-Bi$_2$Se$_3$~\cite{Mishra1997}.  This, combined with challenges to achieve {\em p}-type doped $\alpha$-Bi$_2$Se$_3$, due to natural Se vacancy dominant defect chemistry, make potential thermoelectric applications only of interest at low temperatures~\cite{Hor2009} or in Bi$_2$Te$_x$Se$_{1-x}$ alloys~\cite{Snyder2008}.

There is, however, another polytype of Bi$_2$Se$_3$ with orthorhombic phase, labeled hereafter as o-Bi$_2$Se$_3$, metastable at ambient conditions, that seems to qualify better than its trigonal counterpart for thermoelectric effects~\cite{Fang2020}. This polytype has displayed recent applications in a simultaneous heat flux and temperature acquisition dual-mode sensor~\cite{Kloesel2023}. The Bi$_2$Se$_3$ trigonal-to-orthorhombic phase transition has been shown applications in metal-dielectric-metal metamaterials with tunable negative refractive index in the near-infrared spectral region~\cite{Cao2013}.
Furthermore, o-Bi$_2$Se$_3$ has been proposed as a promising candidate for new photovoltaic interfaces due to an excellent band energy alignment with usual photovoltaic substrates, such as TiO$_2$~\cite{Tumelero2016}.
Phase analysis and knowledge of phase transitions are essential in discovering new applications of materials, sometimes achieved in metastable phase, as is the case of o-Bi$_2$Se$_3$, and sometimes in 2D and nanostructures, such as graphene or wurtzite GaP, respectively. The latter, for example, holds the promise of direct band light emission at 550-570 nm, the so-called green-yellow band, only available in nanowires~\cite{Assali2013}, whereas zincblende GaP is an indirect bandgap semiconductor.

With {\em Pnma} space group, o-Bi$_2$Se$_3$, naturally present as a mineral named Guanajuatite~\cite{Earley1950}, displays quasi one-dimensional atomic ribbons~\cite{Schoenherr2015}, and can be produced as well by quenching it from high-pressure high-temperature growth conditions~\cite{Atabaeva1973,Kang2017} and by electrodeposition of thin films using electrochemical atomic layer epitaxy~\cite{Xiao2009}.

Despite the technological interest, o-Bi$_2$Se$_3$ has received much less attention from the experimental viewpoint compared to its trigonal counterpart, and many fundamental properties such as the electronic bandgap, its direct or indirect nature, and the lattice dynamics properties, such as the phonon dispersion relations, Raman spectra, and their dependence with pressure and temperature, are yet to be fully ascertained.

We report here an investigation of the temperature dependence of the Raman spectrum of o-Bi$_2$Se$_3$ combined with first-principles electronic and lattice dynamics calculations. The role of the anharmonicity, predicted to be larger in orthorhombic than in trigonal Bi$_2$Se$_3$, is also discussed, as well as the strength of the electron-phonon coupling, indicated by the zero-temperature renormalization of the Raman modes.
Moreover, we report here cathodoluminescence data at 83~K that enable us to provide a lower estimate of the experimental electronic bandgap,  
of 0.835~eV, in excellent agreement with the electronic bandstructure calculations.

These results pave the way to further experimental work towards elucidating the role of the lattice potential anharmonicity in the thermal transport, electronic, and optical properties of o-Bi$_2$Se$_3$.

The article is organized the following way:  Section~\ref{sec:methods} describes the sample preparation, the experimental methods of both Raman and cathodoluminescence (CL) measurements, and the details on the {\em ab initio} calculations employed for the analysis.  Section~\ref{sec:results} contributes the observations made in light of the different experiments and calculations, and provides the background for the analysis of the anharmonicity of the Raman modes. This analysis is reported in Sec.~\ref{sec:analysis}. The results are summarized in Sec.~\ref{sec:conclusions} and additional complementary information is provided in the Supplemental Materials (SM)~\cite{SM}.

\section{Methods\label{sec:methods}}
\subsection{Sample preparation}
The commercially available high purity
(99.999\%, Aldrich) $\alpha$-Bi$_2$Se$_3$ powder was used for high-pressure-high-temperature
synthesis of o-Bi$_2$Se$_3$ phase in a toroid type apparatus at $P = 4$~GPa, $T = 673$~K.
A powder sample with a 5~mm diameter and 3~mm height was wrapped in tantalum
foil and placed inside a graphite heater, and then inside a lithographic limestone
container. The heating rate was 20~K~s$^{-1}$. The temperature
holding time under pressure was 60~s, and the cooling rate was 200~K~s$^{-1}$ to quench the structure of the high pressure phase. After cooling
down to room temperature pressure was gradually released obtaining several sintered powder polycrystalline samples.  More details about the synthesis process have been reported in Ref.~\cite{Nadezhda2020}, that also reports information on a new metastable phase with orthorhombic crystal structure, {\em Fdd2} space group, of Bi$_2$Se$_3$. The X-ray diffraction pattern of o-Bi$_2$Se$_3$ is provided in Fig. SM.1 in the Supplemental Materials~\cite{SM}.  
We used an Empyrean Panalytical X-ray diffractometer with Cu K$\alpha$ radiation and an ultrafast high-sensitive X-ray area detector PIXcel3D. The diffractograms were fitted by Rietveld profile-matching using the FULLPROF program~\cite{Carvajal2001}. At each refinement cycle, the fractional coordinates, scale factor, isotropic thermal parameters, profile function, and cell parameters were optimized. Its unit cell parameters are: $a = 11.794$~\AA, $b = 4.1041$~\AA, and $c = 11.574$~\AA. The atomic parameters resulting from the the Rietveld fit are provided in Table~\ref{tabxrd}.

\subsection{Experiments}
Cathodoluminescence (CL) experiments were conducted at 80~K employing
a LEO 1530 Carl Zeiss field-emission scanning electron microscope (SEM) equipped with a MonoCL 2 Gatan UK CL system. Detection was performed with a Peltier-cooled InGaAs detector optimized to work between 900 nm (1.38 eV) and 1800 nm (0.69 eV). An aperture of 60 $\mu$m and an electron beam energy of 20 keV were selected for this experiment. The sample was mounted on a copper holder and contacted with a copper wire to reduce the electron-beam induced charge accumulation on the surface. 
Temperature was controlled by adjusting the flow of liquid nitrogen to
the cold finger in thermal contact with the sample holder, and by activating a heater equipped with a feedback control loop. In this way, the CL spectrum was acquired at 83~K with an accuracy of 1~K at sample stage.

Raman experiments were undertaken in the same sample in the 10~K – 300~K
temperature range, employing a Horiba T64000 triple grating spectrometer and a dye laser of 594 nm excitation line. Spectra were acquired using a liquid nitrogen cooled CCD detector. The spectral resolution was better than 1 cm$^{-1}$, as determined from a Gaussian fit of the spectrum of Ne lines. Power was kept below 1 mW to avoid a temperature-induced phase transformation by laser heating~\cite{Manjon2021}. The sample, displayed in Fig.~\ref{fig1}(a), was kept in a liquid He flow cryostat during the experiment and temperature was read using a PT100 sensor.

\subsection{{\em Ab initio} calculations}

{\em Ab initio} zero-temperature total-energy simulations
were carried out within the framework of density functional theory, DFT~\cite{Hohenberg1964}, as
implemented in the \texttt{Vienna ab initio Simulation Package}~\cite{Kresse1996}, VASP, using projector augmented-wave (PAW) pseudopotentials~\cite{Blochl1994,Kresse1999}.
A plane-wave energy cutoff of 380~eV was employed to ensure accurate converged results. 
The {\em k}-point sampling of the Brillouin zone, BZ, was performed using a 6$\times$14$\times$6 dense grid to ensure
high convergence in the integration over the BZ. 
The exchange-correlation energy was described using the generalized gradient approximation, GGA, with the Perdew-Burke-Ernzerhof (PBE)
functional~\cite{Perdew1998} including the dispersive corrections using the Grimme DFT-D3
method~\cite{Grimme2010}. Spin-orbit coupling was taken into account in the band structure calculations, which were obtained with and without including this term.
The high symmetry path was chosen using the \texttt{SeeK-path} utility~\cite{Hinuma2017}. The unit
cell parameters and the atomic positions were fully relaxed imposing that the forces on the
atoms were less than 0.003 eV/\r{A}, and the deviations of the stress tensors from
a diagonal hydrostatic form were lower than 0.1~GPa. These conditions resulted in $a = 11.852$~\AA, $b = 4.160$~\AA, and $c=11.561$~\AA, in good agreement with the experimental values reported in the literature, e.g., by Ref.~\cite{Tumelero2016b}, of 11.71, 4.11, and 11.43~\AA, respectively, and with the results we obtained by XRD, mentioned above.

Lattice-dynamic calculations of the phonon modes were carried out at the zone centre ($\Gamma$ point) of the BZ with the \texttt{Phonopy} package~\cite{Togo2015}. 
These calculations provide not only the frequency of the normal modes, but also their
symmetry and their polarization vectors. This allows us to identify the
irreducible representations and the character of the phonon modes at the $\Gamma$ point. A 4$\times$4$\times$4 supercell was used in order to obtain the phonon dispersion and the phonon density of states.

\newpage
\section{Results\label{sec:results}}

\begin{figure}[h]
 \includegraphics[width=\linewidth]{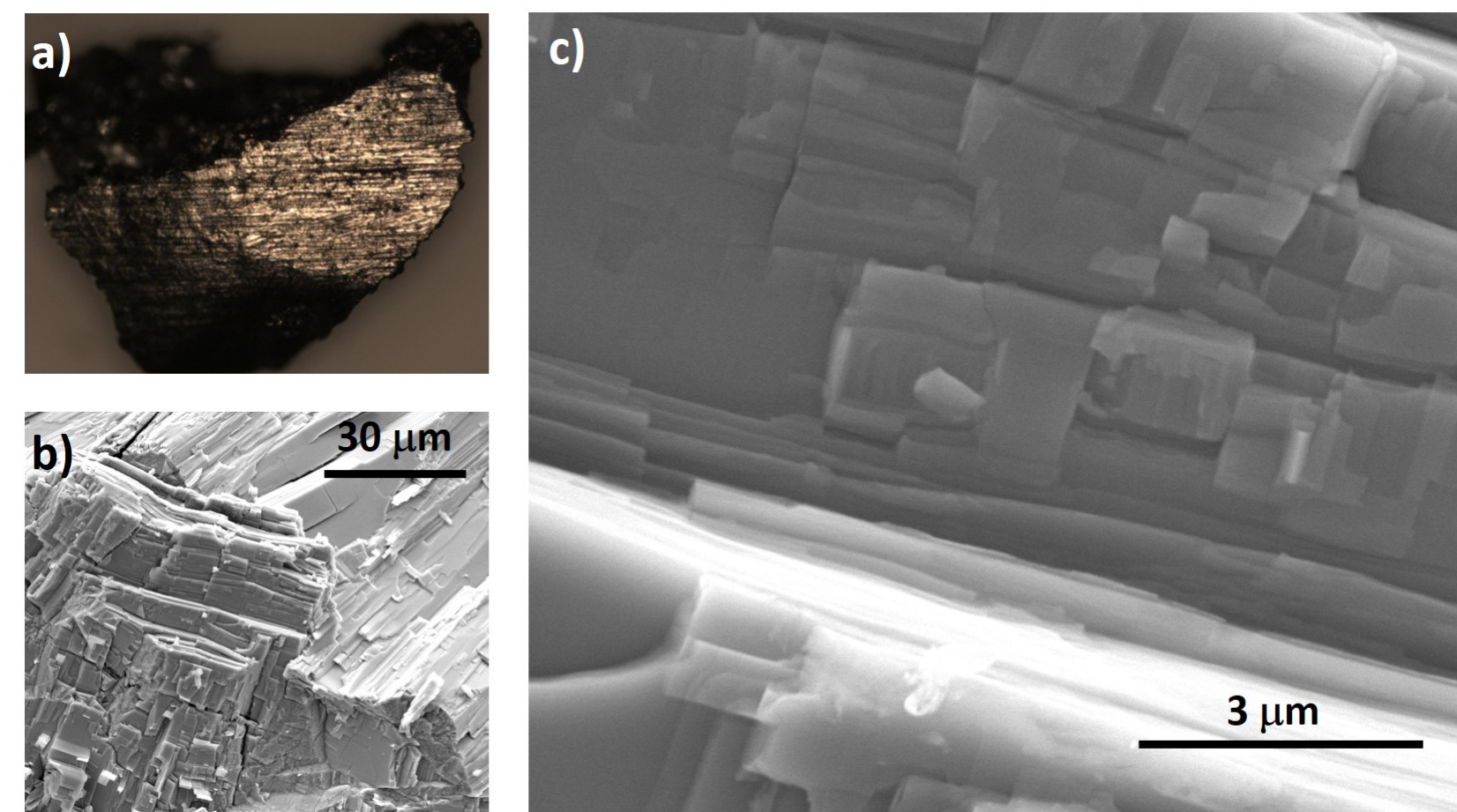}
 \caption{\label{fig1} a) Optical image of the sample under investigation, and b) c) SEM images of o-Bi$_2$Se$_3$ at different magnifications, displaying a stratified structure congruent with an orthorhombic cell.}
 \end{figure}

Figure~\ref{fig1}(a) displays the optical image of the sample under investigation. Figures~\ref{fig1}(b) and (c) show SEM images obtained with 20~keV electron beam energy at several magnifications.   
A stratified structure is observed at the micro- and submicron-scale, with $a$-$c$ planes piled on top of each other. The sample has a shiny metallic-like luster that already indicates an electronic band gap of energies below the visible range. 

\subsection{Electronic bandstructure and bandgap}
\begin{figure}[h]
 \includegraphics[width=0.8\linewidth]{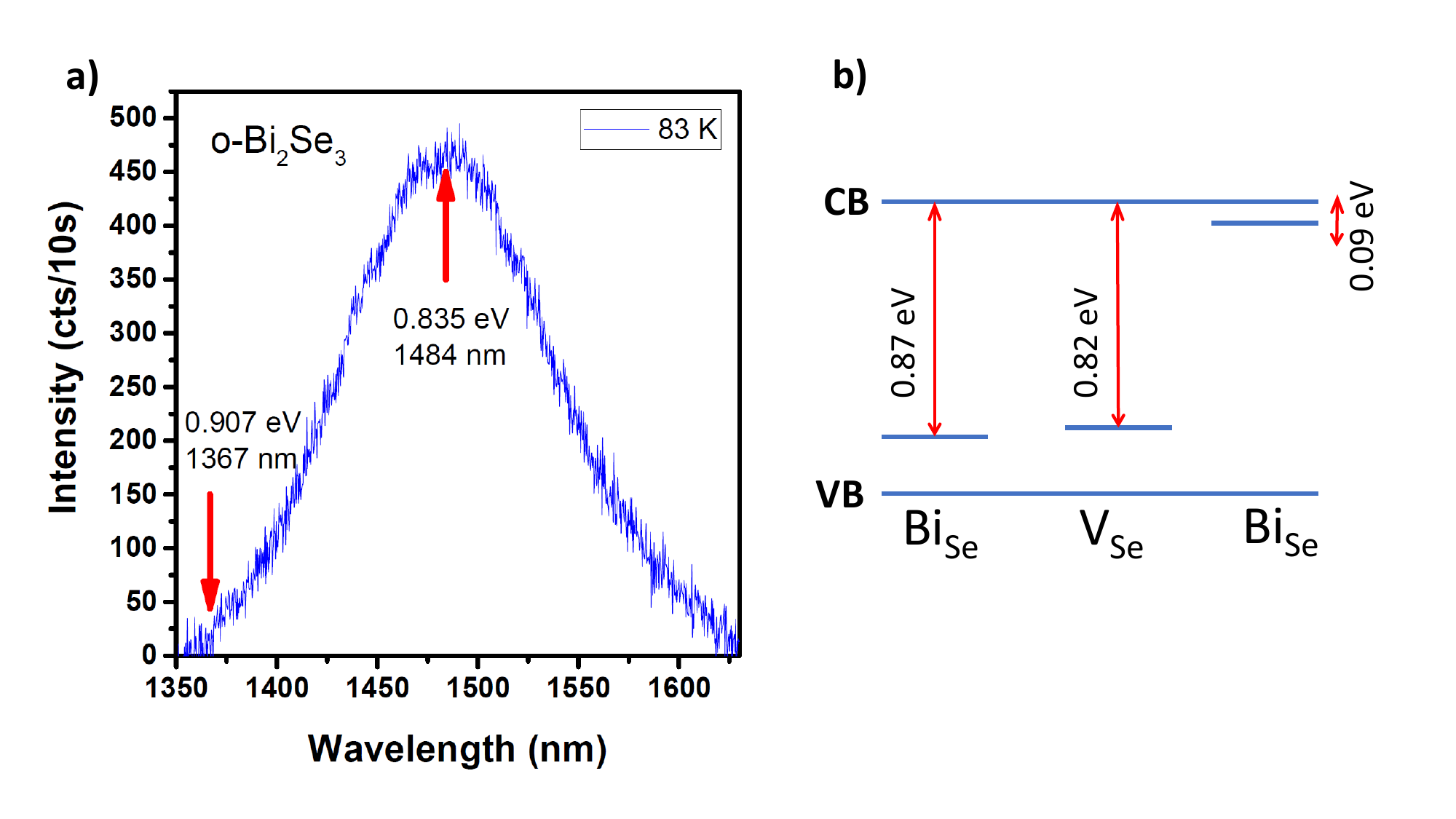}
 \includegraphics[width=0.8\linewidth]{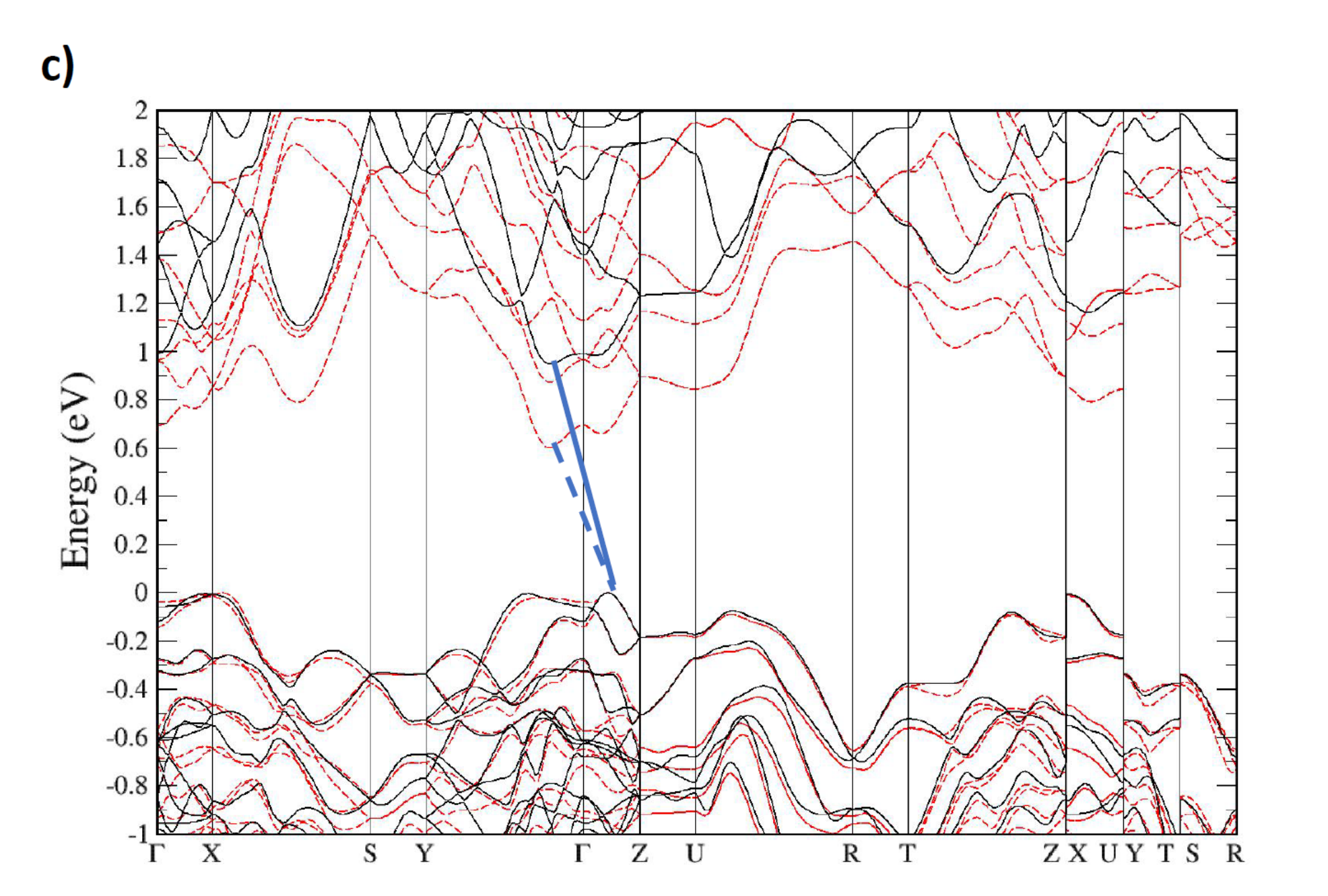}
 \caption{\label{fig2} a) CL spectrum at 83~K of o-Bi$_2$Se$_3$ at the point marked in the SEM image of Fig.~\ref{fig1}(b). b) Schematic bandstructure displaying the top valence band (VB), the bottom conduction band (CB), and three defect energy levels corresponding to donor levels at Bi$_{\rm Se}$ antisites and Se vacancies, and acceptor levels corresponding to the same antisites. Energy values taken from~\cite{Tumelero2016}. c) Calculated {\em ab initio} electronic bandstructures for o-Bi$_2$Se$_3$ along the main symmetry directions of the Brillouin zone with (red dashed lines) and without (black solid lines) including SOC effects. The straight lines display the indirect band gap obtained in both calculations.}
 \end{figure}

The CL spectrum taken at 83~K is shown in Fig.~\ref{fig2}(a), revealing a low-signal broad band peaked at 1484~nm (0.835~eV) and a higher energy edge at 1367~nm (0.907~eV).  Reported optical absorption experiments using near-infrared light on {\em Pnma} Bi$_2$Se$_3$ obtained by electrodeposition revealed a potentially direct energy bandgap, varying between 0.9~eV for 0.5~$\mu$m thick films to 1.25~eV for the thickest films, suggesting their potential for photovoltaic applications~\cite{Tumelero2016b}.
Our estimate for the bandgap energy is consistent with this range.  
In order to shed more light to the nature of this CL band, we calculated the electronic band structure of o-Bi$_2$Se$_3$. This is displayed in Fig.~\ref{fig2}(c) along the main symmetry directions of the BZ, indicating with red dashed and black solid lines the electronic bands obtained  with and without including SOC effects, respectively.

As a general trend, taking into account SOC effects results into lower conduction band energies and a narrowing of the band gap.
These calculations predict an indirect nature of the band gap, located between a maximum of the valence band (VB) along the $\Gamma$-Z direction and a minimum of the conduction band (CB) along the $\Gamma$-Y direction.
Table~\ref{tab1} displays the values of the calculated direct and indirect electronic band gap of orthorhombic Bi$_2$Se$_3$ with and without including SOC effects. For the sake of completeness, our data are compared with those stemming from similar DFT and {\em GW} calculations~\cite{Caracas2005,Sharma2010,Filip2013}. 
One can argue that, due to the common trend of understimate of band gaps of LDA calculations, 
this prediction may be incorrect. However, adding a scissor operator would rigidly shift all electronic bands without affecting the indirect nature of the lowest band gap.
Furthermore, this indirect nature revealed by our calculations, both with and without SOC, is in agreement with predictions made by Ref.~\cite{Filip2013}. 

The calculated band gap energy difference between direct and indirect gaps, however, amounts to 0.1-0.2~eV, which makes it challenging to ascertain the true nature of the gap.  Variations on the lattice parameters results in significant changes in the calculated band gap~\cite{Filip2013}, e.g., up to 0.3~eV, between the values obtained using the experimental lattice constants and those corresponding to DFT fully optimized ones. This fact also prevents the determination of the direct or indirect nature of the band gap based solely on the calculated data.
Cathodoluminescence data show a very weak signal at low temperature, that suddenly vanishes with increasing temperature to 90~K.  This behavior is consistent with defect-related luminescence bands and makes us attribute the band gap to indirect transitions from the VB to the CB. 
If we had a direct band gap, a much larger CL signal would be expected and lasting until higher temperatures. 
After repetition of the experiment several times in different polycrystals of the same batch, only this weak signal at 83~K was observed. 
We therefore conclude that an indirect band gap with some defect bands is the most plausible scenario describing our experimental findings.

Tumelero {\em et al}.~\cite{Tumelero2016} reported a calculated band gap of 1.2~eV and a comprehensive DFT study of the main defect energy levels on o-Bi$_2$Se$_3$, finding donor-type defect levels at 0.84~eV and 0.82~eV from the bottom of the CB corresponding to Bi$_{\rm Se}$ antisite and Se vacancies, respectively. These levels are in agreement with the peak energy of the CL spectrum, located at 0.835 eV.  Figure~\ref{fig2}(b) displays a schematic diagram of the potential transitions between these defect bands and both the VB and the CB to illustrate the possible origin of the features in the CL spectrum. Assuming a bandgap of 0.91~eV according the most accurate {\em GW} calculations with SOC effects~\cite{Filip2013}, an alternative explanation for this peak would be a transition from acceptor states linked to a Bi$_{\rm Se}$ antisite, of 0.09~eV calculated activation energy~\cite{Tumelero2016}. This scenario matches as well with the peak energy of detected CL signal, 0.835~eV, that can be taken as an experimental low energy estimate for the electronic bandgap.

\subsection{Raman phonon frequencies at low temperature}

\begin{figure}[h]
 \includegraphics[width=\linewidth]{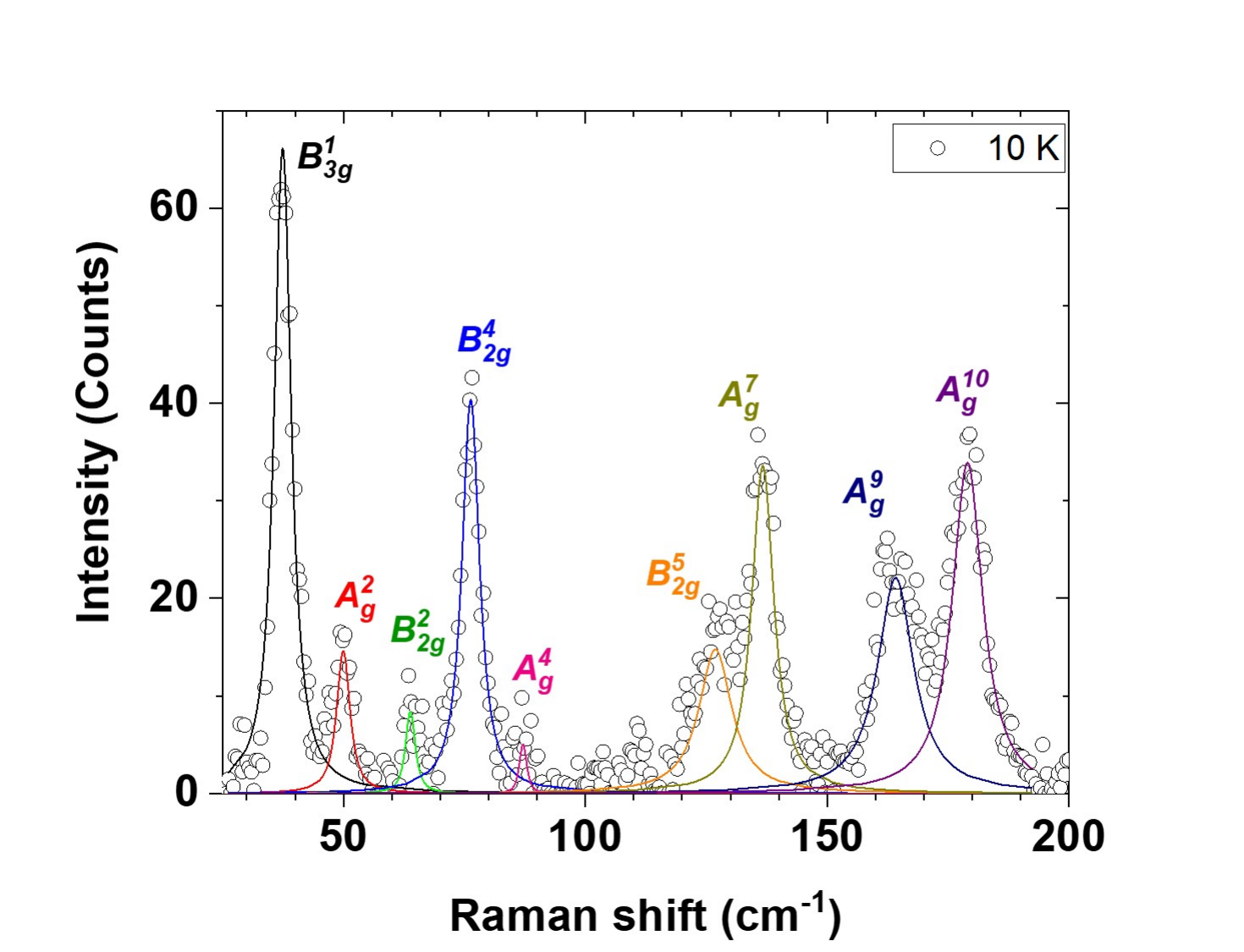}%
 \caption{\label{fig3}Raman spectrum of o-Bi$_2$Se$_3$ at 10~K (symbols). The lines display fits with a Voigt profile taking into account the experimental resolution (approx. 1~cm$^{-1}$). The labels display the assigned
mode symmetry.}
 \end{figure}

Figure~\ref{fig3} shows the Raman spectrum of o-Bi$_2$Se$_3$ obtained at 10~K. Several peaks appear in the spectrum and are attributed to different modes, the symmetry indicated in the figure.   The highest energy mode peaks at about 170~cm$^{-1}$.
To the best of our knowledge, only one work has previously reported the o-Bi$_2$Se$_3$ Raman spectra recently for temperatures between 300~K and 570~K~\cite{Souza2023}.  However, anharmonic effects are expected to be significant at those temperatures given the relatively low energy of the o-Bi$_2$Se$_3$ phonon modes.
A precise determination of these anharmonic effects requires therefore to investigate the temperature dependence of the Raman spectra from 10~K up to room temperature and above.

\subsection{Temperature dependence of the Raman spectrum}
\begin{figure}[h]
 \includegraphics[width=0.7\linewidth]{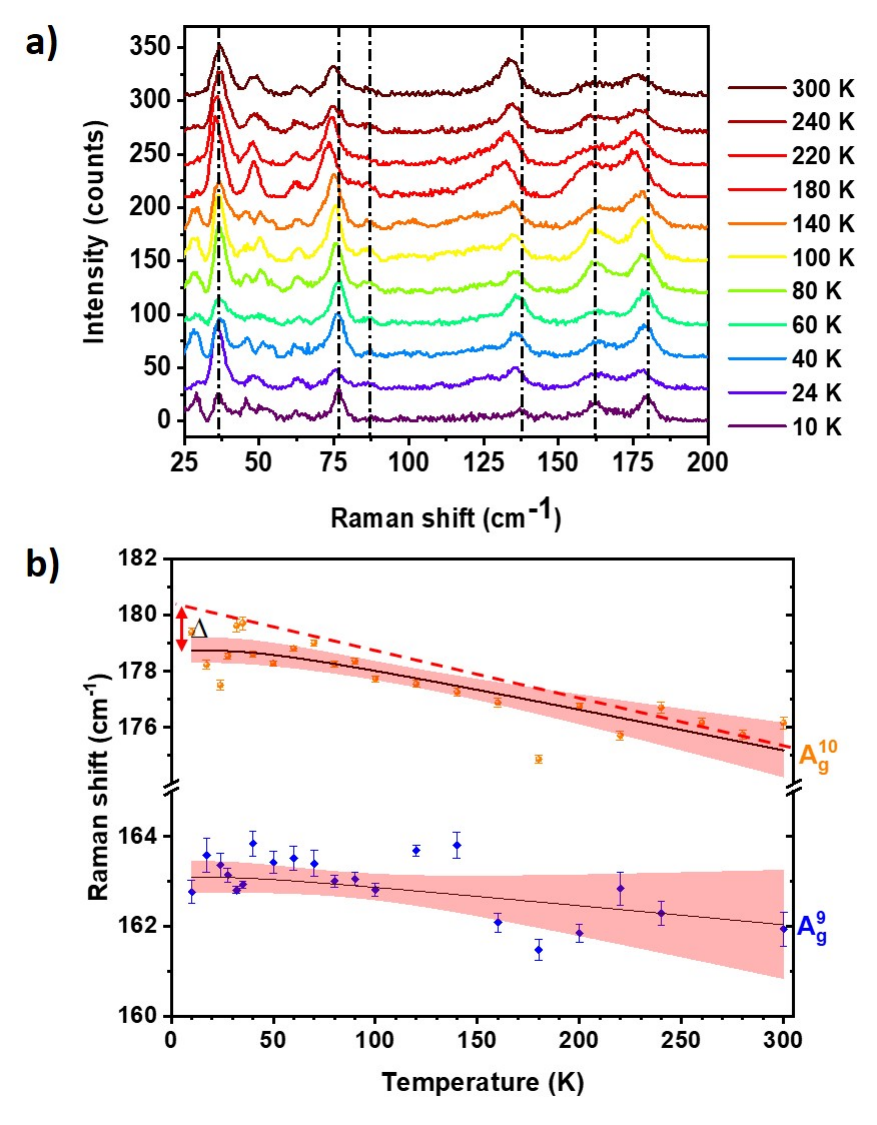}
 \caption{\label{fig4} a)Selected Raman spectra of o-Bi$_2$Se$_3$ as a function of temperature from 10~K to 300~K. The
vertical lines are guides-to-the-eye that display the shift of the most intense modes observed at
10~K upon increase of temperature. b)Temperature dependence of the two highest frequency
modes of o-Bi$_2$Se$_3$ and fit (solid ines) with Eq.~\ref{eqdelta}. The dashed line displays the extrapolation at $T = 0$~K of the high temperature linear behavior for the highest frequency mode, yielding the zero temperature phonon renormalization $\Delta$. Color bands have been used to show the 99\% confidence interval~\cite{Hazra2017}.}
 \end{figure}

Figure~\ref{fig4}(a) plots the Raman spectra at selected values in this 10-300~K temperature range. Some general trends are apparent:  i) Phonon frequencies shift towards lower values with increasing temperature, as expected from the activation of third and high-order anharmonic decay channels, ii) this shift is larger for higher frequency modes, and iii) Raman peaks broaden and Raman intensities tend to decrease with increasing temperature, with a couple of exceptions, also in alignment with an increasing anharmonicity and thereby reduced phonon lifetime. Since the Raman spectra were taken in a polycrystalline sample, we attribute intensity fluctuations and appearance of additional peaks to changes in the laser spot position when the sample temperature is increased. 

A Voigt profile was used to fit the peaks in the Raman spectra. This profile results from the convolution of an intrinsic Lorentzian profile for the phonon excitation and a Gaussian function to take into account the spectrometer resolution function, the latter with a full-width at half-maximum of 1~cm$^{-1}$.

As anharmonic effects are typically more visible for the high energy modes, we display in Fig.~\ref{fig4}(b) the Raman shift of the upper two modes as a function of temperature.  The Raman shift of some of the lowest energy modes is provided as well in Figs.~SM.2,~SM.3, and ~SM.4 in the Supplemental Materials~\cite{SM}.

\newpage
\section{Analysis\label{sec:analysis}}

{\em Pnma}-type Bi$_2$Se$_3$ has an orthorhombic cell with 20 atoms per unit cell, thus yielding a rather complex map of phonon dispersion relations with 60 different phonon branches, similar to Sb$_2$S$_3$ and Sb$_2$Se$_3$~\cite{Fleck2020,Ibanez2016}. Group theory predicts 60 vibrational modes at the Brillouin zone center whose irreducible representations are: $\Gamma = 10 B_{1u} + 5B_{2u} + 10 B_{3u} + 5 B_{1g} + 10 B_{2g} + 5 B_{3g} + 5 A_{u} + 10 A_{g}$, where B$_{1g}$, B$_{2g}$, B$_{3g}$, and A$_{g}$ modes correspond to Raman-active modes, B$_{1u}$, B$_{2u}$, and B$_{3u}$  modes correspond to infrared-active modes, A$_u$ modes are silent, and 1 B$_{1u}$, 1 B$_{2u}$ and 1 B$_{3u}$ modes are acoustic modes. This leads to 30 Raman-active modes and 22 infrared-active modes. 

The lattice dynamics of its trigonal counterpart, with {\em R{$\bar 3$}m} space group, $\alpha$-Bi$_2$Se$_3$, is considerably simpler, as the crystal lattice has 5 atoms in the primitive unit cell. Therefore, only 15 vibrational modes appear at the zone center~\cite{Vilaplana2011,Deshpande2014}. The irreducible representations for the zone-center modes in this case are: $\Gamma = 2A_{1g} + 3A_{2u} + 2E_g + 3E_u$, where A$_{1g}$ and E$_g$ modes are Raman active,  A$_{2u}$ and E$_u$ modes are infrared active,  E$_u$ and E$_g$ are doubly degenerated, and one A$_{2u}$ and one E$_u$ modes correspond to the acoustic phonons.  There are, hence, four Raman-active modes, $2A_{1g}$ and $2E_g$, that are observed at 39~cm$^{-1}$ ($E_g$), 71.5~cm$^{-1}$ ($A_{1g}$),  132.5~cm$^{-1}$ ($E_g$), and 174~cm$^{-1}$ ($A_{1g}$) at room temperature and ambient pressure~\cite{Vilaplana2011,Eddrief2014,Buchenau2020}.
o-Bi$_2$Se$_3$ primitive cell has also 5 atoms, however, the orthorhomic unit cell displays two inequivalent Bi$_2$Se$_3$ ribbons with two unit formulae each, which lead to the need to take 20 atoms into consideration for the lattice dynamics.

The Raman modes of trigonal Bi$_2$Se$_3$ have been subject of intense investigation since the pioneering work by Richter, K{\"o}hler, and Becker in 1977~\cite{Richter1977}, addressing both bulk single crystals~\cite{Zhang2011,Gnezdilov2011,Sharma2021,Rawat2023}, few epitaxial layers~\cite{Shahil2012,Zhao2014,Humlicek2014,Eddrief2014}, thin films~\cite{Niherysh2021}, and nanoplates~\cite{Zhou2018,Kondo2023}. Several studies report the temperature dependence of these modes at different temperature ranges~\cite{Irfan2014,Zhou2018,Gnezdilov2011,Buchenau2020}. Effects of pressure~\cite{Deshpande2014} and magnetic field dependence~\cite{Buchenau2020} have also been investigated in the past.
The lattice dynamics of orthorhombic {\em Pnma} Bi$_2$Se$_3$, on the contrary, has received attention only very recently~\cite{Souza2023}. 

\begin{figure}[h]
 \includegraphics[width=0.8\linewidth]{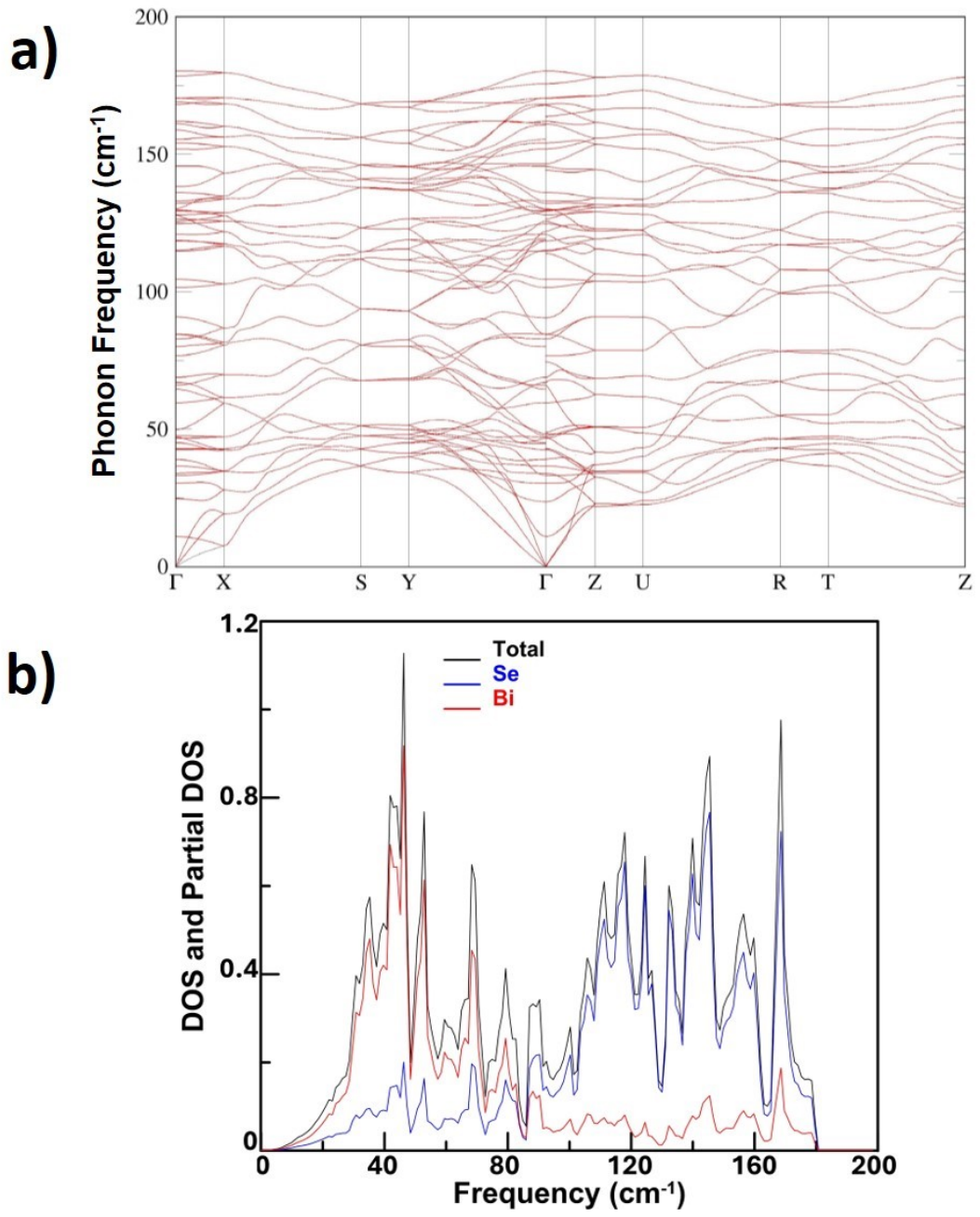}
     \caption{\label{fig:disp} {\em Ab initio}  a)phonon dispersion relations of o-Bi$_2$Se$_3$ along the main symmetry
directions, b)total (black), Bi- (red) and Se-projected (blue) one-phonon density of states.}
 \end{figure}

Figure~\ref{fig:disp}(a) displays the calculated phonon dispersion relations along the main symmetry directions of o-Bi$_2$Se$_3$, whereas Fig.~\ref{fig:disp}(b) shows the total, and Bi- and Se-projected partial one-phonon densities of states (DOS). As expected, the higher frequency modes are dominated by displacements of the Se sublattice. The Van-Hove singularities of the one-phonon DOS are expected to play a role in the peak broadening of the Raman spectra of alloys of o-Bi$_2$Se$_3$, as for example in thermoelectric materials such as in orthorhombic Bi$_2$Te$_x$Se$_{1-x}$, due to disorder-induced breakdown of momentum conservation. The frequencies corresponding to the main Van-Hove singularities are listed in Table~\ref{tabvhove}.

From the 30 Raman-active modes of o-Bi$_2$Se$_3$ only around 10 modes can be identified due to several factors: A large anharmonic broadening, much weaker signal for B$_g$ modes, the spectral resolution of 1~cm$^{-1}$, and the polycrystalline character of the sample under investigation. Furthermore, the spectra acquisition took over an hour per spectrum which led to fluctuations of the 1~$\mu$m diameter laser spot on the sample. This led to changes in Raman intensity for some of the modes as a function of temperature that can be attributed to change in the polycrystal orientation under observation. A similar number of Raman-active modes were previously measured at room conditions for other {\em Pnma}-type related compounds: Sb$_2$Se$_3$, up to 250 cm$^{-1}$~\cite{Fleck2020}, Bi$_2$S$_3$, up to 300 cm$^{-1}$~\cite{Zhao2011}, and Sb$_2$S$_3$, up to 350 cm$^{-1}$~\cite{Sereni2010,Ibanez2016}.

Figure~\ref{fig4}(b) shows a fit to the experimental phonon frequencies with a two-oscillator model assuming anharmonic decay into two modes of equal energy $\omega_1$. This is the so-called Klemens' Ansatz~\cite{Klemens1966}, given by Eq.~\ref{eqdelta}:
\begin{eqnarray}
\label{eqdelta}
    \omega &=& \omega_0 - \Delta \left[1+2n_{\rm BE}\left(\omega_1,T\right)\right] \\
    &=& \omega_0 - \Delta \left[ 1 +\frac{2}{e^{\frac{\hbar\omega_1}{k_B T}}-1}\right] \nonumber
\end{eqnarray}
where $\omega_0$ is the unrenormalized zero-temperature phonon energy, $\Delta$ is the energy renormalization at 0~K, $n_{\rm BE}$ is the Bose-Einstein function, and $\hbar$  and $k_B$ are Planck's  and Boltzmann's constants, respectively.  $\omega_1$ is a phonon frequency lower than $\omega_0$.

Energy conservation must be fulfilled for anharmonic decay.  This is reflected in the rule $\omega_0 = \omega_1 + \omega_2$ for the allowed transitions in the calculation of the second-order anharmonic contributions to the phonon linewidth.  
However, in the calculation of the same contribution to the phonon frequency shift, a Kramers-Kronig transformation must be applied, mixing decay channels that do not require for energy conservation. This is explicitely displayed in the exact equations corresponding to the third-order anharmonic contributions to the phonon shift and linewidth, provided by Eqs. 2.14a and 2.15a in Ref.~\cite{Balkanski1983} for Silicon, and it also holds for binary and multinary compounds~\cite{Serrano2012}. Note the delta functions in Eqs. 2.14 and 2.15~\cite{Balkanski1983}, responsible for energy conservation, appear only in the expressions of the anharmonic contributions to the linewidth. 
Hence, $\omega_1$ may be taken as an adjustable independent parameter.

In order to simplify the number of fitted parameters, we have imposed $\omega_1 = \omega_0 /2$. The quality of the fit can be assessed using 99\% confidence intervals as defined in Ref.~\cite{Hazra2017}. 
These are displayed in Fig.~\ref{fig4}(b) and further figures using a colored band. 
Unfortunately, the quality of the data prevents from extracting additional information on other decay channels, e.g. such as the electron-phonon contribution in related materials with topological insulators (Ref.~\cite{Singh2022}).

Since {\em ab initio} phonon frequencies are usually obtained within the harmonic approximation, comparison between experimental and calculated Raman frequencies is more accurate and appropriate using the values obtained for $\omega_0$, rather than the experimental $\omega$ values at low temperature. Investigation of anharmonicity using {\em ab initio} simulations requires much more complex calculations similar to those reported in the pioneering work of Debernardi {\em et al}. in Ref.~\cite{Debernardi1995}, and it is beyond the scope of this work.

In the high temperature limit, $n_{\rm BE} (\omega,T) \approx \hbar\omega/(k_B T)$.  From this limit, together with Eq.~\ref{eqdelta}, we can calculate the slope of the temperature dependence of the Raman phonons at the high temperature range.  Given the relatively low Raman frequencies of o-Bi$_2$Se$_3$, one can safely assume that at room temperature we are in this high $T$ regime, and then
\begin{equation}
    \label{eq2}
    \frac{\partial\omega}{\partial T} \approx -4\frac{k_B \Delta}{\hbar\omega_0}.
\end{equation}

Table~\ref{tab2} displays the calculated Raman frequencies both with and without taking into account SOC effects, with the values of $\omega_0$ and $\Delta$ derived from our experimental results, the Raman frequencies obtained at 300~K, and those reported in Ref.~\cite{Souza2023}. We also report here the high temperature limit of the temperature derivative of the Raman phonon frequencies, as calculated with Eq.~\ref{eq2}. Another estimate for the temperature derivative at room temperature can be obtained by using the approximation $\frac{\partial\omega}{\partial T} \approx (\omega_{\rm 300 K} - \omega_0)/300$.  In this case, much lower values are obtained, e.g., for the highest frequency modes an average temperature derivative of $-0.012$ and $-0.014$~cm$^{-1}$ K$^{-1}$, similar to those reported for trigonal Bi$_2$Se$_3$ modes (between $-0.012$ and $-0.024$~cm$^{-1}$ K$^{-1}$)~\cite{Kim2012,Yang2012,Zhou2018}.
The values of $\Delta$ give information about the zero-temperature renormalization of the phonon frequency, and are typically larger the more anharmonic the material.   Note that in Table~\ref{tab2} only the largest energy mode displays significant renormalization, 2.1~cm$^{-1}$, the value being comparable to those reported for isotopic CuI, 2.0-2.6~cm$^{-1}$, a prototypical anharmonic semiconductor with similar highest Raman frequency~\cite{Serrano2012}. 

\begin{figure}[h]
 \includegraphics[width=\linewidth]{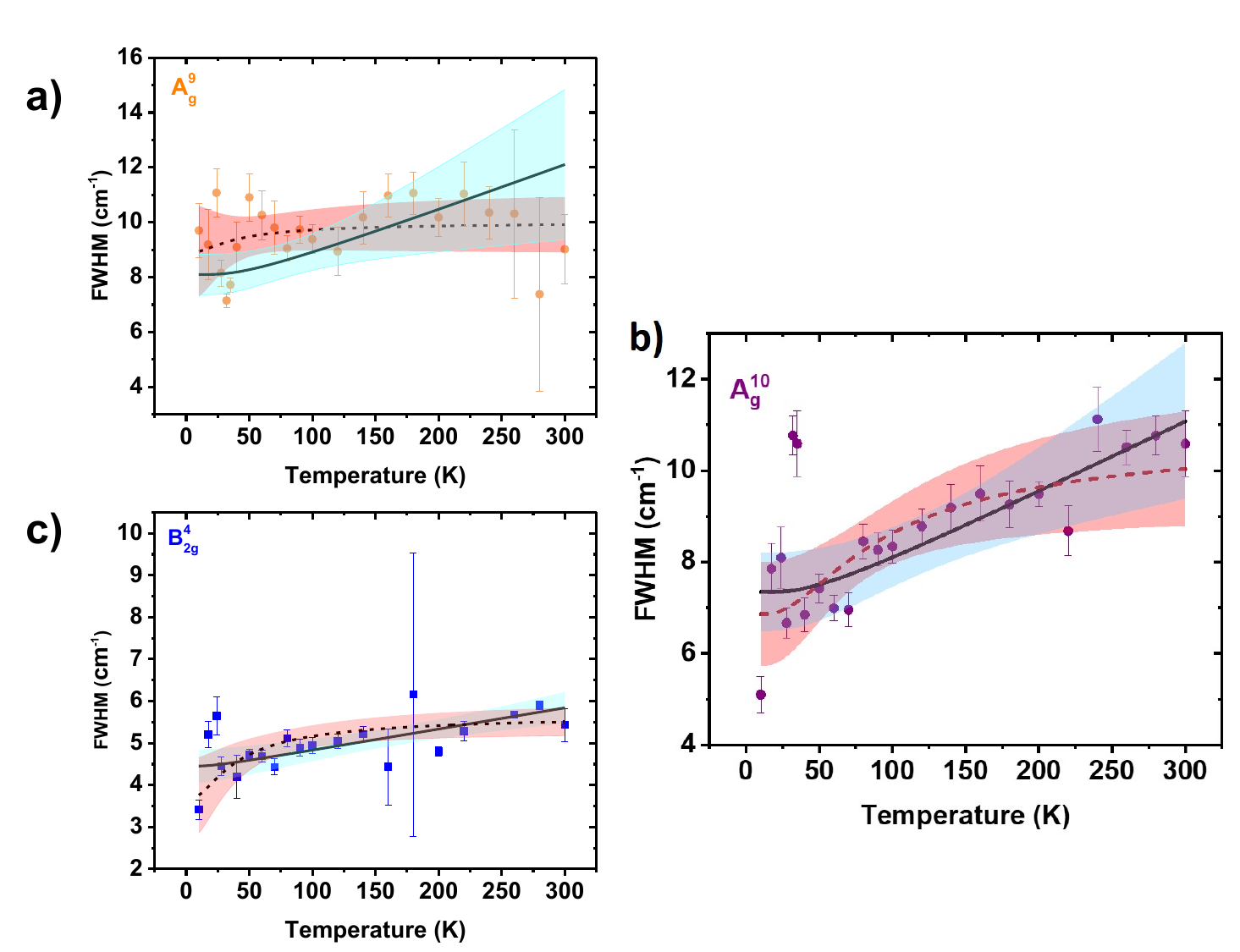}
\caption{\label{fig6} Temperature dependence of the intrinsic phonon FWHM of the o-Bi$_2$Se$_3$ Raman modes corresponding to a) $\omega_0 = 164$~cm$^{-1}$ ($A_g^9$), b) $\omega_0 = 180$~cm$^{-1}$ ($A_g^{10}$), and c) $\omega_0 = 76$~cm$^{-1}$ ($B_{2g}^4$). The solid line displays fits with Eq.~\ref{eqgamma2klem}, whereas the dashed lines indicates a fit with a decay channel of two phonons of different energies~\cite{Menendez1984}, using Eq.~\ref{eqgamma2men}. A much better fit is obtained in Fig.~\ref{fig6}(a) and (c) with the latter equation. However, in Fig.~\ref{fig6}(b) both fits display a similar quality revealing the need to be cautious with these fits. Color bands have been used to show the 99 confidence interval~\cite{Hazra2017}. }
 \end{figure}

We can, therefore, conclude that the anharmonic effects are significant in this material. However, the anharmonicity seems to be lower than that exhibited by trigonal Bi$_2$Se$_3$. The latter presents both larger linewidths of the Raman peaks and an overshooting at 50~K of the higher A$_{1g}$ Raman frequency as a function of temperature, indicating phonon interactions with the electronic system beyond the expected anharmonic decay~\cite{Buchenau2020}.
 Figure~\ref{fig6}(a) displays the phonon linewidth for the $A_g^9$ mode, with $\omega_0 = 164.8$~cm$^{\rm -1}$.  Significant data scattering is observed and attributed to a varying mixture of mode intensities due to the fluctuations of the laser spot in the sample at different temperatures. Unfortunately, for o-Bi$_2$Se$_3$
 the large number of Raman modes in a relatively small energy range  challenges a thorough analysis of the anharmonicity through direct investigation of the phonon linewidths as a function of temperature. 
Despite this overlap, we were able to evaluate the temperature dependence of the linewidth of other two modes, $B_{2g}^4$ and $A_g^{10}$, corresponding to bare frequencies $\omega_0 = 76$~cm$^{-1}$ and 180~cm$^{-1}$, displayed in Fig.~\ref{fig6}(c) and (b), respectively. 

The anharmonic contribution to the phonon linewidth, $\Gamma_{\rm anh}$ (FWHM), is related to the phonon frequency shift due to anharmonic decay, $\Delta$, by a Kramers-Kronig transformation, since both correspond to the real and imaginary parts of the phonon self-energy $\Sigma(\omega)$~\cite{Balkanski1983}:
\begin{eqnarray}
\label{eqgamma}
\Sigma(\omega) &=& \Delta(\omega) - i\frac{\Gamma_{\rm anh}(\omega)}{2}{\rm , where\ using\ the\ Klemens'\ Ansatz} \nonumber \\
    \Gamma_{\rm anh} &=& \Gamma_1 \left[1+2n_{\rm BE}\left(\omega_1,T\right)\right] \\
    &=& \Gamma_1 \left[ 1 +\frac{2}{e^{\frac{\hbar\omega_1}{k_B T}}-1}\right].\nonumber
\end{eqnarray}
In order to describe properly the experimental phonon linewidths we need to introduce an additional broadening, 
$\Gamma_0$, that takes into account crystalline disorder effects and other phonon scattering sources such as defects. This term may account as well for fluctuations due to experimental errors.

If we modify Eq.~\ref{eqgamma} to introduce this broadening, we can fit the linewidths with the following equation~\cite{Ulrich2022}:
\begin{equation}
\label{eqgamma2klem}
    \Gamma = \Gamma_0 +\Gamma_1 \left[1+2n_{\rm BE}\left(\omega_1,T\right)\right].
\end{equation}
Equation~\ref{eqgamma2klem} provides a good fit in some cases, as in Fig.~\ref{fig6}(a), dotted line, where disorder and defect effects dominate.  However, as Menendez and Cardona showed for crystalline Si~\cite{Menendez1984}, in other cases Klemens' Ansatz falls short to explain the observed temperature dependence of Raman phonon linewidths, due to an excessive weight of overtone modes and lack of consideration of other two-phonon combinations as anharmonic decay channels, respecting momentum and energy conservation rules, such as 
$q_1 = -q_2$ and $\omega_0 = \omega_1 + \omega_2$.  

The generalized expression for Eq.~\ref{eqgamma2klem} including decay into two modes with different frequencies is given then by:
\begin{equation}
\label{eqgamma2men}
    \Gamma = \Gamma_0 +\Gamma_1 \left[1+n_{\rm BE}\left(\omega_1,T\right) +n_{\rm BE}\left(\omega_2,T\right)\right],
\end{equation}
and it has been used in the literature to describe properly the temperature dependence of Raman phonon linewidths for other diatomic semiconductors, such as CuI~\cite{Serrano2012}.

In Figs.~\ref{fig6}(b) and (c), a sudden increase in linewidth is observed below 100~K, followed by a more moderate increase at higher temperatures.  This peculiar behavior may have two distinct causes: i) Either it is a signature of a strong anharmonicity for o-Bi$_2$Se$_3$, or ii) it is an effect of a large disorder component in the linewidth added to a moderate anharmonic term.
In the first case, the observed trend requires the use of Eq.~\ref{eqgamma2men} to be accounted for properly, whereas the second case can be explained employing the simpler Klemens' expression, Eq.~\ref{eqgamma2klem}.
Confidence intervals have been used with fits employing each equation to compare the quality of both hypotheses, and are displayed with red and blue bands, respectively.

In order to apply a fit with Eq.~\ref{eqgamma2men}, we first need to validate that there are either flat bands or bands with opposite slope in the phonon dispersion that can account for anharmonic decay channels consisting of combination of two phonons with different energy, $\omega_1$ and $\omega_2$, with sum equal to the Raman energy of the mode under analysis. A more thorough approach consists of calculating the two-phonon DOS corresponding to sum modes and searching for Van Hove singularities at the Raman energies~\cite{Serrano2012,Serrano2003}.  Due to constraints of the program we used to obtain our {\em ab initio} phonon energies, this option was not available.

Visual inspection of the phonon dispersion relations reveals combinations of flat bands of acoustic and optic modes that can yield more likely decay channels for the three Raman modes mentioned above. Thus, e.g., for $A_{g}^{10}$, $\omega_0 = 180.5$~cm$^{-1}$, we take $\omega_1 = 67.5$~cm$^{-1}$ and $\omega_2 = 113$~cm$^{-1}$, that correspond to flat phonon dispersions along the U-R high symmetry direction.  Similarly, for $B_{2g}^4$, $\omega_0 = 76$~cm$^{-1}$, we can assign $\omega_1 = 25$~cm$^{-1}$ and $\omega_2 = 51$~cm$^{-1}$, corresponding to the flat phonon dispersion along the Z-U high symmetry direction. For $A_g^9$, $\omega_0 = 164$~cm$^{-1}$, a similar argument can be employed using a combination of modes located at 55~cm$^{-1}$ and 109~cm$^{-1}$, i.e., corresponding to two maxima of the one-phonon DOS due to flat dispersions along S-Y direction, among others.
Note that these choices correspond roughly to $\omega_1=0.33\ \omega_0$ and 
$\omega_2=0.66\ \omega_0$, similar to the values used in Ref.~\cite{Menendez1984} to obtain an accurate description of the Raman phonon linewidths for silicon.

The linewidth values plotted in Fig.~\ref{fig6} have been obtained after deconvolution of the Raman peak, in a similar way as described in Ref.~\cite{Buchenau2020}. Table~\ref{tabgamma} displays the values obtained for the fitting parameters of Eqs.~\ref{eqgamma2klem} and~\ref{eqgamma2men} to the displayed modes: $B_{2g}^4$, $A_{g}^{9}$, and $A_{g}^{10}$.
Color bands have been added to show the 99\% confidence interval and ascertain the quality of both fits.
Whereas both equations yield relatively good fits for the $B_{2g}^4$ mode, and similar values for defect- and anharmonic-attributed broadening for $A_{g}^{9}$, they seem to disagree in the description of the  $A_{g}^{10}$ phonon linewidth.  Klemens' equation tends to give more weight to the contribution of disorder and defects in the linewidth, while Menendez and Cardona's equation reveals a larger anharmonic effect.  This calls for caution when using these equations to infer the extent of either contribution to the broadening.  
Since Eq.~\ref{eqgamma2men} takes into account implicitly both overtones and combinations of phonons, a more accurate description that that of Klemens', when both fits have a similar quality, we recommend to choose Menendez and Cardona's formalism~\cite{Menendez1984}.

Following this line of argumentation, the higher value of $\Gamma_1$, 4.0(9)~cm$^{-1}$, observed for the higher frequency mode, $A_{g}^{10}$, compared to that obtained for the $B_{2g}^4$ mode ($\Gamma_1 = 2.0(4)$~cm$^{-1}$), agrees with the larger anharmonic effects in the phonon frequency observed also for this mode. This can be attributed to a larger value of the two-phonon density of states, since there is also a critical point in the phonon bandstructure at both 36~cm$^{-1}$ and 144~cm$^{-1}$ that may yield another suitable decay channel for the $A_{g}^{10}$ mode.

Further modes have been analyzed with no relevant changes to the conclusions of the manuscript, and their main results are shared in the Supplemental Materials~\cite{SM}.  It is worth noting that different modes exhibit different temperature dependence both in the linewidth and in the Raman shift due to varying decay channels and anharmonic coupling matrix elements~\cite{Balkanski1983,Menendez1984}.

\section{Conclusions\label{sec:conclusions}}
We have reported here the temperature dependence of the Raman spectrum of orthorhombic, {\em Pnma}, Bi$_2$Se$_3$ in the 10~K -- 300~K temperature range, and analyzed the Raman shifts using a two-oscillator Klemens-like model.
The Raman shifts at 10~K are in agreement with the predictions we obtained using linear-response {\em ab initio} calculations. An anomalous anharmonic behavior is observed in the temperature dependence of the phonon linewidths for at least two of the Raman modes, displaying a two-slope trend with increasing temperature. This can be attributed to a combination of large defect concentration and significant anharmonic effects.  More experimental work with additional samples should be done to ascertain the weight of each contribution.
We have also observed by cathodoluminescence a lower limit of 0.835~eV for the electronic bandgap of o-Bi$_2$Se$_3$. This band gap seems to have an indirect nature, given the low intensity observed in the experimental data, and also in agreement with both our calculations and those previously reported using the {\em GW} approximation.
These results shed light to some of the fundamental properties of orthorhombic Bi$_2$Se$_3$, a metastable material of interest for applications in thermoelectrics.

\begin{acknowledgments}
This work is financed by the Spanish Ministerio de Ciencia e Innovaci\'on and the Agencia Estatal de Investigaci\'on MCIN/AEI/10.13039/501100011033 as part of the project MALTA Consolider Team Network (RED2022-134388-T), and I+D+i projects PID2019-106383GB-42/43, PID2021-126046OB-C22, TED2021-130786B-I00 also funded by MTED, and PID2022-138076NB-C42/C44 co-financed by EU FEDER funds. It is also funded by project PROMETEO CIPROM/2021/075 (GREENMAT), financed by the Generalitat Valenciana, and by Generalitat Valenciana through project MFA/2022/025 (ARCANGEL), and also forms part of the Advanced Materials programme supported by the Spanish MCIN with funding from European Union NextGenerationEU (PRTR-C17.I1). I.M-M. also acknowledges financial support through a PhD contract by Junta de Castilla y Leon, Spain.

\end{acknowledgments}

\bibliography{bi2se3}

\newpage

\begin{table}[h]
\begin{ruledtabular}
\caption{o-Bi$_2$Se$_3$ atomic parameters obtained from the refinement of XRD spectrum given in Fig. SM.1 in the Supplemental Materials~\cite{SM}. The experimental lattice parameters obtained from the fit were $a = 11.7936$~\AA, $b = 4.1041$~\AA, and $c = 11.5724$~\AA.}
\label{tabxrd}
\begin{tabular}{|ccddd|}
Atom & Site & x & y& z \\\hline
Bi & Bi1 & 0.01332 & 0.25000 & 0.32957 \\
Bi & Bi2 & 0.34599 & 0.25000 & 0.54081 \\
Se & Se1 & 0.05123 & 0.25000 & 0.86597 \\
Se & Se2 & 0.88154 & 0.25000 & 0.55105 \\
Se & Se3 & 0.22378 & 0.25000 & 0.19475 \\
\end{tabular}
\end{ruledtabular}
\end{table}

\begin{table}[h]
  \caption{Calculated electronic bandgap energy, in meV, for {\em Pnma}-type Bi$_2$Se$_3$ obtained with and without including spin-orbit coupling (SOC), and peak energy observed in CL spectrum at 83~K. Values from other DFT and from {\em GW} calculations were taken from $^{\rm a}$Ref.~\cite{Caracas2005}, $^{\rm b}$Ref.~\cite{Sharma2010}, and $^{\rm c}$Ref.~\cite{Filip2013}.}
    \label{tab1}
    \begin{tabular}{|l|c|c|}
    \hline
         & Direct gap & Indirect gap  \\\hline
        Exp. (CL data) & & 835(10)\\\hline
        Without SOC & 974 & 948 \\
        With SOC & 624 & 601  \\
        Other DFT & 900$^{\rm a}$, 1100$^{\rm b}$, 990$^{\rm c}$& 860$^{\rm c}$  \\
        GW with SOC & 910$^{\rm c}$ &  \\
        \hline
    \end{tabular}
\end{table}

\newpage
\begin{table}[h]
\begin{ruledtabular}
\begin{tabular}{|ccc|ccc|}
$\omega$ & {\rm DOS} & Assignment & $\omega$& {\rm DOS} & Assignment \\
({\rm cm}$^{-1}$) &    &            & ({\rm cm}$^{-1}$) &    &           \\\hline
31 & 0.40 & $\Gamma$X    & 111 & 0.61 & SY,RT \\
35 & 0.58 & $\Gamma$X,ZU & 118 & 0.72 & RT    \\
43 & 0.80 & RT           & 124 & 0.67 & UR    \\
46 & 1.13 & $\Gamma$X,SY & 132 & 0.60 & ZUR   \\
53 & 0.77 & UR           & 143 & 0.73 & $\Gamma$Z \\
68 & 0.65 & SY           & 145 & 0.89 & $\Gamma$Y  \\
79 & 0.41 & SY,ZU        & 156 & 0.54 & $\Gamma$X  \\
90 & 0.34 & SY,$\Gamma$Z & 160 & 0.48 & $\Gamma$Z  \\
100& 0.28 & RT           & 169 & 0.98 & $\Gamma$X  \\
107& 0.41 & ZU           & 178 & 0.17 & $\Gamma$X  \\ 
\end{tabular}
\end{ruledtabular}
\caption{Calculated frequencies of the main Van-Hove singularities of the one-phonon DOS and the corresponding assignments in light of the {\em ab initio} phonon dispersion relations.}
\label{tabvhove}
\end{table}

\newpage
\begin{table*}[h]
\smallskip
\begin{ruledtabular}
\scalebox{0.7}{
\begin{tabular}{c Sc d d d d d d c}
Mode & Symmetry &\multicolumn{2}{c}{Theory, $\omega_0$}&
\multicolumn{4}{c}{Experiment} &  Ref.~\cite{Souza2023}\\
 & & \mc{With SO} & \mc{No SO} & \mc{$\omega_0$} & \mc{$\Delta$} & \mc{$\omega$(300~K)} & \mc{$\frac{\partial\omega}{\partial T}$($\frac{{\rm cm}^{-1}}{\rm K}$) } & \mc{$\omega$(300~K)} \\ \hline
1 & B$_{2g}^1$ & 30.7 & 31.1 & 28.85(15) &0.04(2) & 28.4^{\rm 280~K} & -0.004 & \\
2 & B$_{1g}^1$ & 30.7 & 31.7 &   &  & & &\\
3 & A$_{g}^1$ & 32.7 & 34.2 &   &  & & &\\
4 & B$_{3g}^1$ & 34.2 & 35.4 & 36.7(2)  &-0.05(3)  &37.3 & 0.004 & \\
5 & B$_{1g}^2$ & 44.9 & 46.4 & 45.4(9)  &-0.2(2)  &48.4 & 0.012&\\
6 & B$_{3g}^2$ & 45.7 & 47.3 &   &  & & &\\
7 & A$_{g}^2$ &  45.7 & 47.0 & 50.5(6)  & 0.0(2) & 50.6^{\rm 140~K}& 0.000&\\
8 & A$_{g}^3$ &  58.3 & 59.9 & & & & &\\
9 & B$_{2g}^2$ & 64.6 & 67.2 & 63.37(15) & 0.06(4) & 63.0 & -0.003 &\\
10 & B$_{2g}^3$ & 67.1& 69.0 &  &&  &  &\\
11 & B$_{2g}^4$ & 80.4 & 82.6 & 76.6(3) & 0.46(9)& 74.9& -0.017 &\\
12 & A$_{g}^4$  & 82.4 & 84.6 & 88.2(10)& 1.7(4) & 78.3& -0.054 &\\
13 & B$_{1g}^3$ & 98.6 & 101.3 & & & & &\\
14 & B$_{3g}^3$ & 102.1& 104.5 & & & & &\\
15 & B$_{2g}^5$ & 111.9& 114.0 & & & & &109.0 \\
16 & A$_{g}^5$  & 112.1& 115.0 & & & & &\\
17 & A$_{g}^6$  & 115.7& 117.3 & & & & &\\
18 & B$_{2g}^6$ & 118.7& 121.7 & & & & &\\
19 & B$_{3g}^4$ & 120.4& 124.9 & & & & &\\
20 & B$_{1g}^4$ & 124.1& 128.4 & 127.3(12)& -0.1(7)& 129.5 & 0.002 & 126.6\\
21 & A$_{g}^7$  & 128.1& 130.1 & 137.6(4)& 1.4(2)& 134.0 & -0.028  & \\
22 & B$_{1g}^5$ & 128.7& 132.8 & & & & &\\
23 & B$_{3g}^5$ & 131.4& 135.4 & & & & &\\
24 & B$_{2g}^7$ & 142.5& 145.7 & & & & & 141.1\\
25 & B$_{2g}^8$ & 148.7& 151.7 & & & & &\\
26 & A$_{g}^8$  & 151.0& 153.7 & & & & &\\
27 & A$_{g}^9$  & 158.2& 161.2 & 164.8(4)& 1.2(3)& 161.1& -0.020 & 158.7\\
28 & B$_{2g}^9$ & 163.4& 168.5 & & & & &\\
29 & A$_{g}^{10}$& 167.0& 170.6& 180.5(4)& 2.1(3)& 176.2& -0.032 & 173.0\\
30 & B$_{2g}^{10}$&167.4& 170.3& & & & &\\ 
\end{tabular}
}
\end{ruledtabular}
\caption{Calculated and measured Raman frequencies for o-Bi$_2$Se$_3$ for the different symmetry modes. The unrenormalized phonon frequencies, $\omega_0$ and their renormalization at 0~K, $\Delta$, extracted from the experimental data using Eq.~\ref{eqdelta}, are also displayed here. Raman frequencies are given in cm$^{-1}$, and temperature derivatives were calculated from the experimental data using Eq.~\ref{eq2}. Experimental frequencies are given at 300~K except for some modes that were not observed at room temperature. In those cases, the temperature at which they are observed is indicated with a superscript.}
\label{tab2}
\end{table*}

\newpage

\begin{table}[h]
\begin{ruledtabular}
\caption{Inhomogeneous ($\Gamma_0$) and anharmonic ($\Gamma_1$) contributions to the phonon linewidth of the Raman modes displayed in Fig.~\ref{fig6}, obtained using 
Eqs.~\ref{eqgamma2klem} and~\ref{eqgamma2men}, taking into account the error bars in both fits. Note that, for A$_g^{10}$,
 fits with both equations yield nearly the same quality, reduced 
Chi-square values equal to 7.04 and 7.63, respectively, leading to 
different relative weight of both contributions.}
\label{tabgamma}
\begin{tabular}{|ldd|}
Mode & \multicolumn{1}{c}{$\Gamma_0$~(cm$^{-1}$)} & \multicolumn{1}{c}{$\Gamma_1$~(cm$^{-1}$)}  \\\hline
B$_{2g}^4$ (Eq.~\ref{eqgamma2klem})& 4.3(2) & 0.14(2)  \\
B$_{2g}^4$ (Eq.~\ref{eqgamma2men})& 1.7(7) & 2.0(4)  \\\hline
A$_g^9$    (Eq.~\ref{eqgamma2klem})& 7.0(5) & 1.1(3) \\
A$_g^9$    (Eq.~\ref{eqgamma2men})& 7.8(15) & 1.1(9) \\\hline
A$_g^{10}$ (Eq.~\ref{eqgamma2klem}) & 6.3(5) & 1.0(2) \\
A$_g^{10}$ (Eq.~\ref{eqgamma2men}) & 2.8(12) & 4.0(9) \\
\end{tabular}
\end{ruledtabular}
\end{table}

\newpage

\end{document}